\documentclass[aps,twocolumn,superscriptaddress]{revtex4}
\usepackage{latexsym}
\usepackage{graphicx}
\usepackage{amsfonts,amssymb}

\begin{document}

\title{Phase transitions in Ising model induced by weight redistribution on weighted regular networks}
\author {Menghui Li}
\affiliation{Department of Systems Science, School of Management,
\\Center for Complexity Research, Beijing Normal University, Beijing
100875, P.R.China.}
 \affiliation{Temasek Laboratories, National University of
Singapore, 117508, Singapore}

\author {Ying Fan}
\affiliation{Department of Systems Science, School of Management,
\\Center for Complexity Research, Beijing Normal University, Beijing
100875, P.R.China.}

\author {Jinshan Wu}
\affiliation{Department of Physics \& Astronomy, University of
British Columbia, Vancouver, B.C. Canada, V6T 1Z1.}

\author {Zengru Di \footnote{Author for correspondence: zdi@bnu.edu.cn}}
\affiliation{Department of Systems Science, School of Management,
\\Center for Complexity Research, Beijing Normal University, Beijing
100875, P.R.China.}

\begin{abstract}
In order to investigate the role of the weight in weighted networks,
the collective behavior of the Ising system on weighted regular
networks is studied by numerical simulation. In our model, the
coupling strength between spins is inversely proportional to the
corresponding weighted shortest distance. Disordering link weights
can effectively affect the process of phase transition even though
the underlying binary topological structure remains unchanged.
Specifically, based on regular networks with homogeneous weights
initially, randomly disordering link weights will change the
critical temperature of phase transition. The results suggest that
the redistribution of link weights may provide an additional
approach to optimize the dynamical behaviors of the system.
\\
\\
Pacs: 89.75.Hc, 05.70.Fh, 64.60.Fr
\\

\end{abstract}

\maketitle

\section{Introduction}
Complex networks are widely used to describe the interaction
structure of many complex systems. Since the Watts-Strogatz (WS)
model\cite{D.J.Watts} was proposed, the structure, function, and
evolution of complex networks are investigated extensively. In
particular, many kinds of dynamical processes evolving on networks
are studied, e.g., the spread of infectious disease\cite{D.J.Watts},
fast response and coherence of Hodgkin-Huxley
neurons\cite{coherence}, percolation\cite{Resilience}, security of
system (Cascade-based attacks)\cite{grid}, phase
transitions\cite{A.Barrat,Andrzej}, focusing on the influence of
topological structure of the underlying network on the dynamical
behaviors.  All these studies try to understand the role of
topology, usually by disordering the links. For instance, based on
regular networks, one can construct small-world networks by rewiring
or adding links randomly. For disorder of the rewiring type, the
nearest neighbor links are rewired with the rewiring probability
$p_r$ to form random shortcut links\cite{D.J.Watts}. For disorder of
the adding type, without removing local links, random shortcut links
are added with the probability $p_a$\cite{Jespersen,Mousson} or with
the probability $P(l)\sim l^{-\delta}$, where $l$ is the Euclidean
distance between two vertices\cite{P.Sen}. Both types display phase
transitions from regular networks to small-world networks by varying
the single parameter $p_r$ or $p_a$ representing disorder.

The behaviors of dynamics on small-world networks are usually
compared to those on regular networks to understand the role taken
by network topologies. In this paper, we focus on the phase
transition of the Ising models on weighted networks\cite{Andrzej}.

The phase transition of the Ising model on small-world networks was
investigated extensively in the past years. Barrat and
Weigt\cite{A.Barrat} and Gitterman\cite{M.Gitterman} studied the
crossover from one-dimensional to mean-field behavior for the
ferromagnetic Ising model, which presents a phase transition of
mean-field type for any value of the rewiring probability $p_{r}>0$.
Later, Herrero\cite{C.P.Herrero} investigated the ferromagnetic
transition of the Ising model on small-world networks generated by
rewiring two-dimensional and three-dimensional lattices. In the
thermodynamic limit, the phase transition has a mean-field character
for any finite value of the rewiring probability $p_{r}$. In
 Ref.\cite{D.Jeong}, D. Jeong et al studied the Ising model on
small-world networks with the coupling strength $J_{ij}$ decaying
algebraically with the distance $r$, i.e., $J_{ij}(r)\propto
r_{ij}^{-\alpha}$, where $r_{ij}$ is the geometrical distance
between two vertices $i$ and $j$ on the underlying one-dimensional
lattice. Das and Sen\cite{quench} studied the quenching dynamics of
the ferromagnetic Ising model on densely connected small-world
networks at zero temperature. Recently, Chatterjee and
Sen\cite{Arnab} investigated the critical behavior of the Ising
model on one-dimensional networks, where long-range bonds were taken
into consideration with the probability $P(l)\sim l^{-\delta}$ if
distances $l>1$.  In addition, the self-averaging properties of the
Ising model on networks were also investigated recently, such as the
magnetization $M$ and the susceptibility $\chi$\cite{S.Roy}.

Most of the above studies of Ising models on networks have mainly
concentrated on disordering the topological structure of regular
lattices. In fact, link weights, directly related to the coupling
strength of spins, can also affect the phase transition of the Ising
model. For instance, the topology-dependent coupling strength, e.g.
$J_{ij}\propto (k_ik_j)^{-\mu}$, had remarkable influence on the
phase transition of the Ising model\cite{PRE_74_036108}. In the case
that the coupling strength is directly related to the distance
$d_{ij}$ between vertices, e.g. $J_{ij} \propto d_{ij}^{-\alpha}$,
where $d_{ij}$ depends on link weights on the path, an interesting
question is: How does randomizing link weights affect the process of
phase transition of the Ising model?

In many realistic networks besides the connecting structure
(described by links), interaction intensity (described by link
weights) is also an important property of networks, and usually
plays an important role as to the dynamical processes evolving on
the networks. For example, the number of passengers or flights
between any two airports in airport networks\cite{W.Li,Barrat1}, the
closeness of any two scientists in scientific collaboration
networks\cite{Barrat1,Newman1,Newman2,Li}, and the reaction rates in
metabolic networks\cite{Nature} are all crucial to characterize the
corresponding systems. Recently more and more studies in complex
networks focus on the effect of link weights on dynamics, such as
epidemic spread\cite{ZhouT}, transportation\cite{Goh,KIGoh,Vragovi},
percolation\cite{Guanliang},
synchronization\cite{Chavez,CZhou,CSZhou}, functional
organization\cite{function} and so on.

For weighted networks, disordering link weights provides another way
to adjust the structure and to optimize the dynamical behaviors of
networks. In our previous papers, we introduced one method to
disturb the weight-link correspondence\cite{MLi} and another
mechanism to redistribute link weights\cite{Menghui}, and
investigated their influences on properties of networks. With
randomly redistributed link weights, the average path length
decreases, while the average clustering coefficient
increases\cite{Menghui}. This indicates that random redistribution
of link weights may induce small-world phenomena. In addition,
redistributing link weights can enhance synchronizability of chaotic
maps on weighted regular networks\cite{Menghui,Daqing}. In this
paper, we consider Ising spins on the vertices of weighted regular
networks and investigate the effects of weights redistribution on
the phase transition.

This paper is organized as follows. In Section \ref{Disorder}, we
will briefly introduce the method of disordering link weights. Then,
in Section \ref{Ising}, we will give the results of the Ising model
on weighted regular networks. We find that disordering link weights
has significant influence on the critical temperature of phase
transition, which reflects the important role taken by the link
weights in networks. Finally, in Section \ref{conclud}, some
concluding remarks are given.

\section{The method of disordering link weights}\label{Disorder}
In weighted networks, link weights can be represented as measures of
dissimilarity or similarity. For dissimilarity weight $w$, e.g., the
distance between two airports, the distance between two vertices,
which are connected by a third vertex and two links (with
dissimilarity weights $w_{1}$ and $w_{2}$ respectively), is defined
as $d=w_{1}+w_{2}$. For similarity weight $\tilde{w}$, e.g. the
number of cooperations between any two scientists in scientific
collaboration networks or the coupling strength between oscillators,
the similarity distance between two vertices, which are connected by
a third vertex and two links (with similarity weights
$\tilde{w_{1}}$ and $\tilde{w_{2}}$ respectively), is defined as
$\tilde{w}=1/({\frac{1}{\tilde{w_{1}}}+\frac{1}{\tilde{w_{2}}}})$,
which is smaller than both $\tilde{w_{1}}$ and $\tilde{w_{2}}$.
Distance can be defined as the inverse of similarity weight, i.e.
$d=1/\tilde{w}$. In the following discussion, the dissimilarity
weight is chosen as $w_{ij}\in\left[1,\infty\right)$, and
consequently the similarity weight $\tilde{w}_{ij} = 1/w_{ij}$ is in
$\left(0,1\right]$\cite{Menghui}.

\begin{figure}
\centering
\includegraphics[width=0.8\linewidth]{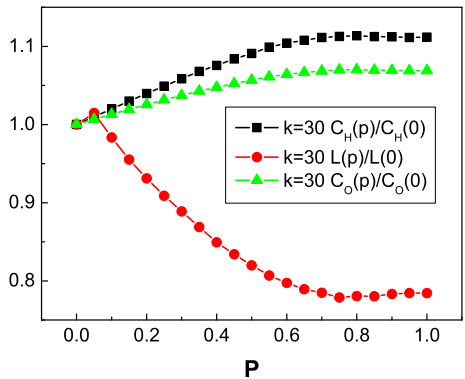}
\caption{Without rewiring links, characteristic path length
$L(P)/L(0)$ and clustering coefficient $C(P)/C(0)$ for the family of
randomly weight redistributed networks ($N=300, k=60, w=10$).
$C_{H}$ and $C_{O}$ are defined in Eq.(\ref{newcc}) and
Eq.(\ref{newOnnela}) respectively. The $x$-axis is the probability
of redistributing link weights, and the $y$-axis is the value of
$L(P)/L(0)$ and $C(P)/C(0)$, where $L(0)$ and $C(0)$ are the values
of initial uniform regular networks. All results are averaged over
$20$ random realizations of the disordering process and the relative
standard deviation is less than $2\%$.} \label{SWNWeight}
\end{figure}

Our initial setup is a ring lattice with $N$ vertices. Every vertex
links to $k$ nearest neighbors. Each link has the same dissimilarity
weight, e.g. $w=10$, which corresponds to the distance. All
connections are undirected. We assume that there is a minimum unit
of weight, e.g., $\Delta w=1$. The procedures of disordering link
weights are as follows:
\begin{enumerate}
\item Every unit of weight in the original lattice is removed with
probability $P$ from the original link, and transferred to a link
randomly chosen over the whole lattice.

\item Step 1 is repeated until each unit of weight in the original
lattice has been tried once. The reallocated weights will not be
considered again.

\item If the unit of weight being selected is the last unit
left on that link, it will not be moved. This is to avoid
disconnecting the link so that the topology remains unchanged.
\end{enumerate}
Without changing the binary structure, above procedures allow us to
adjust the network with uniform link weight ($P=0$) to the one with
poisson weight distribution ($P=1$). We are able to discuss and
compare quantities such as distance and clustering coefficients for
networks before and after disordering link weights. In our
calculation, $\tilde{w}\in (0,1]$, we revise the definition of
weighted clustering coefficient\cite{Onnela,BJKim} as follows(see
details in Ref.\cite{Menghui}),
\begin{equation}
C^w_H{(i)} =
\frac{\sum_{j,k}\tilde{w}_{ij}\tilde{w}_{jk}\tilde{w}_{ki}}{\sum_{j,k}\tilde{w}_{ij}\tilde{w}_{ki}}
\label{newcc}
\end{equation}
and
\begin{equation}
C^w_O(i)=\frac{2}{k_i(k_i-1)}\sum_{j,k}(\tilde{w}_{ij}\tilde{w}_{jk}\tilde{w}_{ki})^{1/3}.
\label{newOnnela}
\end{equation}
With disordering link weights, the average path length $L(P)/L(0)$
decreases, while the average clustering coefficient $C(P)/C(0)$
increases(as shown in Fig.\ref{SWNWeight}). This demonstrates that
besides rewiring links, disordering link  weights also leads to
small-world phenomena. This provides a potential approach to
optimize the dynamical behaviors of system.

In the following section, we will investigate the Ising models on
weighted regular networks with the coupling strength $J_{ij}$
decaying algebraically with distance $d_{ij}$, where $d_{ij}$
represents the shortest distance between two vertices $i$ and $j$.
By investigating system behaviors in the intermediate region $0 < P
< 1$, we can study the effect of disordering link weights on the
phase transition of Ising models.

\section{Ising Model on Weighted Regular Networks}\label{Ising}
We perform Monte Carlo (MC) simulations with the heat bath algorithm
at various values of $P$. The Hamiltonian for an Ising model on
weighted regular networks is given by
\begin{equation}
H=-\frac{1}{2}\sum _{i\neq j} J_{ij} \sigma _i  \sigma
_j,\label{Hamiltonian}
\end{equation}
where $\sigma _i(=\pm 1)$  is the Ising spin on vertex $i$. The
distance-dependent interaction $J_{ij}$ reads
\begin{equation}
J_{ij}=d_{ij}^{-\alpha},\label{Hamiltonian}
\end{equation}
where $d_{ij}$ is the shortest path length (rather than the
geometrical distance) between two vertices $i$ and $j$. The role of
$\alpha$ has been investigated in
Ref.\cite{PRL_29_917,PRL_89_025703}, indicating that (i)
mean-field-type critical behavior for $1<\alpha<1.5$; (ii)
continuously varying critical exponents for $1.5<\alpha<2$; (iii)
for $\alpha =2$ a hybrid transition, with a jump in magnetization
but continuous energy is expected, and the short-range interaction
regime with no phase transition for $\alpha>2$.
 In this paper, we mainly focus on the influence of redistributing
link weights on the phase transition of the Ising model, so we only
investigate the mean field type, e.g. $\alpha=1$ for simplification.
Although the value $\alpha =1$ is somewhat singular, as there are
stability issues (e.g. in the thermodynamic limit the ground state
energy per site is infinite), in our simulations these singularities
do not show up, and we assume  that the model behaves like a regular
model with  mean-field exponents, which corresponds to our numerical
findings.

For each realization of networks, every vertex is given a spin
($\sigma_i=1$ or $\sigma_i=-1$) randomly with equal probability, so
the magnetization is nearly zero at the beginning (paramagnetic
phase). Then, the system evolves according to  the metropolis
algorithm.

\begin{figure*}[th]
\includegraphics[width=0.4\linewidth]{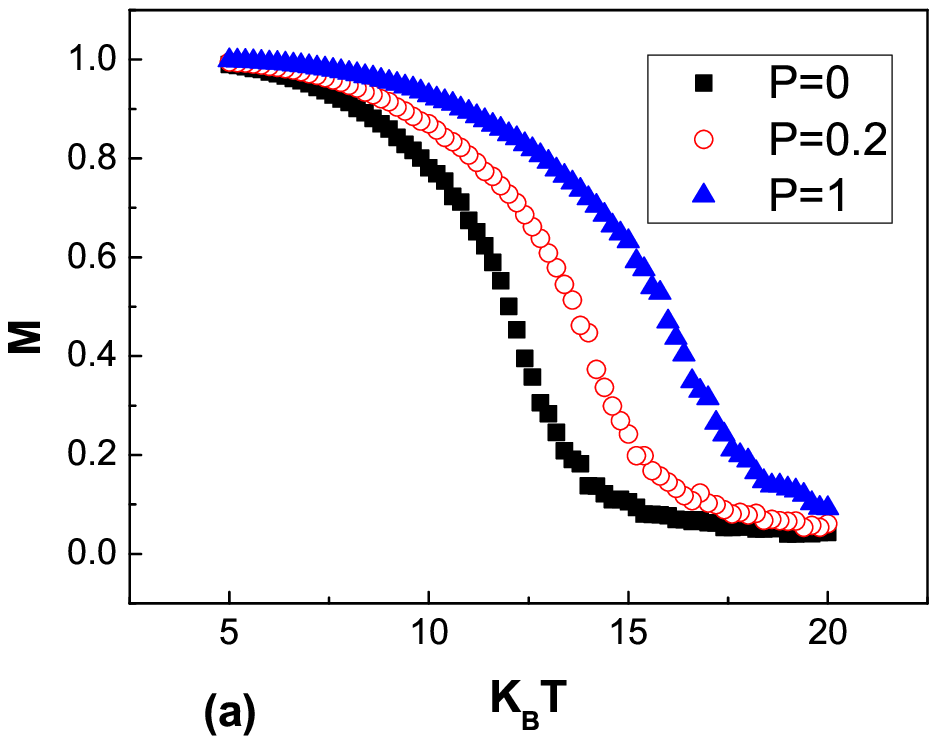}\includegraphics[width=0.4\linewidth]{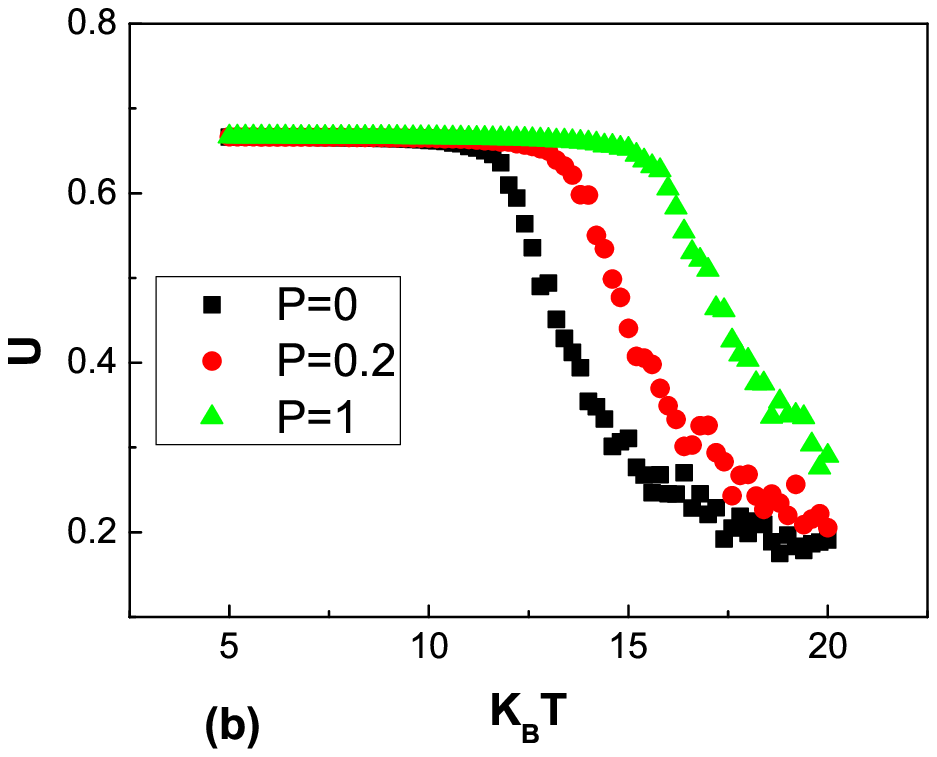}
\includegraphics[width=0.4\linewidth]{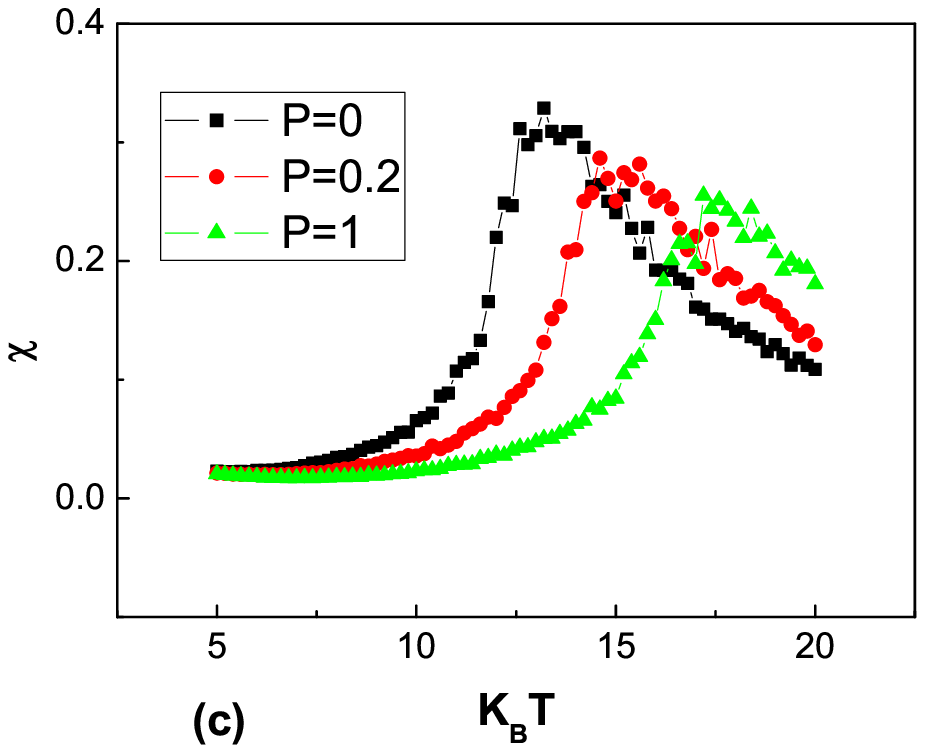}\includegraphics[width=0.4\linewidth]{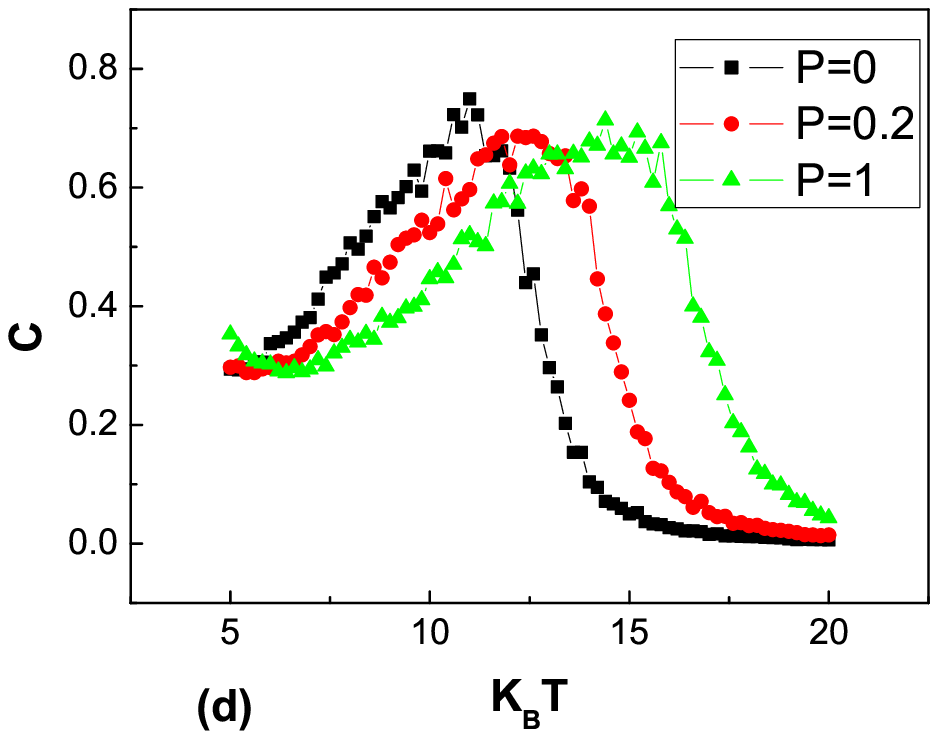}
\caption{Quantities of Ising model  versus temperature $K_BT$ on
weighted regular networks with interaction $J_{ij}=d_{ij}^{-1}$ for
$P=0, 0.2, 1$. (a) Magnetization per spin $M$,
 (b) Binder¡¯s cumulant $U$, (c) Susceptibility per spin $\chi$, (d)
Specific heat capacity per site $C$.} \label{critial}
\end{figure*}

The following quantities are computed on weighted regular networks
after the system reaches equilibrium:
\begin{enumerate}
\item  \emph{Magnetization per spin}
\begin{equation}
M=\sum \sigma_i/N.
\end{equation}
\item  \emph{Binder¡¯s fourth-order cumulant}
\begin{equation}
U=1-\frac{\langle m^4\rangle}{3\langle m^2\rangle^2}.
\end{equation}

\item  \emph{Susceptibility per spin} is calculated from the
fluctuation of the order parameter:
\begin{equation}
\chi= \frac{N}{K_BT}(\langle m^2\rangle - \langle m\rangle^2),
\end{equation}
where $K_B$ is the Boltzmann¡¯s constant.

\item \emph{Specific heat capacity per spin} is obtained from the
energy fluctuations at a given temperature

\begin{equation}
C=\frac{\langle H^2\rangle - \langle H\rangle^2}{N(K_BT)^2}.
\end{equation}
\end{enumerate}
Here $\langle \cdot \cdot \cdot \rangle$ denotes the thermal
average, taken over $3000$ Monte Carlo steps after discarding $3000$
Monte Carlo steps for equilibration at each temperature. All results
are averaged over $20$ different network realizations and over $10$
random realizations of the redistribution process, respectively.

The simulations are made on a ring of typically $N=300$ vertices
with $k=60$ neighbors per vertex, and last until the system reaches
a stationary state, where the magnetization fluctuates around the
average value. When $P=0$, link weights are all homogeneous, while
for $P\neq 0$ link weights are heterogeneous. After disordering link
weights, the average shortest path length decreases
clearly(Fig.\ref{SWNWeight}). Figure \ref{critial} presents the
quantities as the functions of temperature on weighted regular
networks with $P=0, P=0.2$ and $P=1.0$. For different $P$, the
curves of magnetization are significantly different. With the
increasing of $P$, the critical temperature moves to higher values
(as shown in Fig.\ref{critial} (a)). Correspondingly, other curves
are also quite distinct for different $P$, such as Binder's
fourth-order cumulant, Susceptibility per spin and Specific heat
capacity per spin. The appearance of the susceptibility peak in
Fig.\ref{critial} (c) as well as the specific heat peak in
Fig.\ref{critial} (d) unanimously suggests that the phase transition
emerges at a finite temperature. At temperatures where $\chi$ and
$C$ display peaks, the fluctuations are very large. These give some
information that critical phenomena could be observed at the peak
temperatures. It can be seen clearly (Fig.\ref{critial} (c)) that
when the disordering probability ($P$) increases,  the peak
temperature of the susceptibility $\chi$ keeps increasing while the
peak value of $\chi$ keeps decreasing.

\begin{figure}
\center
\includegraphics[width=0.8\linewidth]{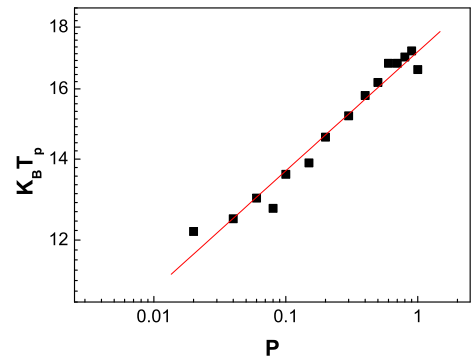}
\caption{Peak temperature $T_P$ versus the probability $P$ of
disordering link weights on the log-log plane. The line is the power
fit.} \label{peak}
\end{figure}

\begin{figure}
\centering \includegraphics[width=0.8\linewidth]{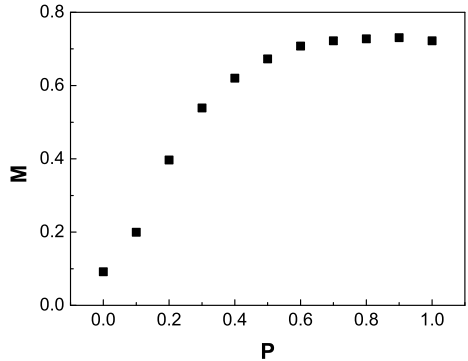}
\caption{Magnetization $M$ versus the probability $P$ of disordering
link weights when $K_BT=14$.} \label{tem}
\end{figure}

In numerical simulations, it is very difficult to determine the
critical temperature $T_c$. Instead of $T_c$, we will observe the
peak temperature $T_p$ of the susceptibility. Figure \ref{peak}
gives the dependence of the peak temperature on the disordering
probability $P$, indicating that the peak temperature increases with
the disordering probability $P$ raising. The peak temperature as a
function of the disordering probability $P$ is given by a power law
\begin{equation}
T_p \sim P^{\beta},
\end{equation}
where $\beta \simeq 0.10$. This indicates that disordering link
weights has diminishing marginal effects. With the raising of
disordering probability $P$, it has smaller effect on the shift of
the peak temperature.

At a fixed temperature, the value of order parameter $M$ increases
as the disordering probability $P$ raises (as shown in Fig.
\ref{tem}). Fortunately, the magnetization grows rapidly around
$T_p$. This implies that there is no doubt about the onset of the
ordering.

\begin{figure}
 \includegraphics[width=0.8\linewidth]{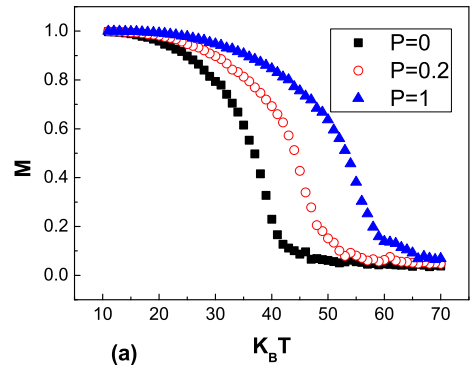}  \includegraphics[width=0.8\linewidth]{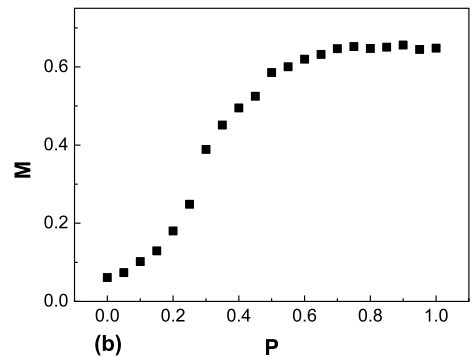}
\caption{Results in larger regular network with $N$=1000 and
$k$=200. (a) Magnetization per spin $M$. (b)Magnetization versus the
probability $P$ of disordering link weights when $K_BT=50$.}
\label{bigger}
\end{figure}

Since the system is too small, the phase transition at fixed
temperature is not obvious. We also make simulations on a larger
ring of typically $N=1000$ vertices with $k=200$ neighbors per
vertex. Figure \ref{bigger} gives the curves of Magnetization $M$ as
functions of the temperature and the disordering probability at
fixed temperature. The systems of different sizes are ordered at
different temperature $T$ as the ground state energy per spin
diverges with the size $N$ increasing(as shown in
Fig.\ref{critial}(a) and Fig.\ref{bigger}(a)). This is not a key
issue in our model as we only care about the influence of
disordering link weights on the critical temperature of the same
size systems. Fortunately, the systems with same parameters are
exactly ordered at the same temperature. Obviously, the system is in
the paramagnetic phase ($\langle M\rangle\approx 0$) at $P=0$, and
it is in the ordered, ferromagnetic phase ($|\langle M\rangle| >
0.5$) at $P=1$(as shown in Fig.\ref{bigger}(b)).

In all the simulations, the connecting structures of networks remain
unchanged. The only factor, which can affect the behavior of phase
transition, is the redistribution of link weights. This demonstrates
that link weight plays an important role in the function of a
network system.

We restrict the system size to $N$ in our studies, because a large
number of configurations are required to get accurate results. The
final results are the averages over $200$ simulations, including
$10$ initial configurations of spins on the same network and $20$
different networks generated with the same parameters. As the
networks are densely connected, it consumes a lot of CPU time to
search and update shortest paths. We have to limit our studies on
relative small system size.

\section{Concluding Remarks}\label{conclud}
As a measure of interaction strength, link weight is believed to be
an important factor in networks. It gives more properties of the
network besides topological structure, and provides an additional
dimension to adjust network properties. For weighted networks,
besides changing topology, redistributing link weights is an
important way to optimize the dynamical behaviors of networks.

In this paper, we investigate the phase transition of the Ising
model on weighted regular networks with interaction $J(d)\sim
d^{-1}$. We have performed Monte Carlo simulations on networks with
different probability $P$ of disordering link weights. The results
show that disordering link weights can affect the critical
temperature of phase transition. At fixed temperature, the order
parameter $M$ attains the value $M\simeq0$ for the paramagnetic
phase at $P=0$ and a value $M>0$ for the ordered, ferromagnetic
phase at $P=1$.

Although the effect of disordering weights is not so significant as
that of changing topology, it is an important supplement way, and
provide an additional approach to adjust the dynamical behaviors of
the system.

This paper mainly focuses on the Ising model on weighted regular
networks, where $J_{ij}=d_{ij}^{-\alpha}$ and $\alpha=1$. Of course,
other value of $\alpha$ may be more useful to depict the effect of
disordering link weights. Other dynamical process can also be
investigated on weighted networks by similar approach. For further
studies, the more interesting problem is to seek the best matching
pattern between link weight and topological structure to optimize
the dynamical behaviors of the network. This would be very useful
for designing networks and optimizing the dynamical behavior of the
system.

\section*{Acknowledgments}
 Author M.H. Li wants to thank Xiaofeng Gong(NUS) for his reading
and helpful comments. This work is partially supported by $985$
Projet, NSFC under the grant No. 70771011 and No. 60974084.

\bibliography{apssamp}

\end{document}